\documentclass[aps,prb,reprint,showpacs,superscriptaddress]{revtex4-2}
\usepackage{blindtext}
\usepackage{epsfig,amssymb,amsmath,subfigure,hyperref  } 
\usepackage{siunitx}
\usepackage{mathrsfs}
\usepackage{array}
\usepackage{lipsum}
\usepackage{color}
\usepackage{multirow}
\usepackage{tabularx}
\usepackage{ragged2e} 
\usepackage{bm}

\usepackage{graphicx}
\usepackage{times}
\usepackage[english]{babel}

\begin{document}
\title{Time crystalline solitons and their stochastic dynamics in a driven-dissipative $\phi^4$ model}

\author{Xingdong Luo}
	\email{luoxingdong@jssnu.edu.cn}
	\affiliation{School of Physics and Electronic Information, Jiangsu Second Normal University, Nanjing 211200, China} 
    
\author{Zhizhen Chen}
	\email{ChenZhiZhen@sjtu.edu.cn}
	\affiliation{Wilczek Quantum Center and Key Laboratory of Artificial Structures and Quantum Control, School of Physics and Astronomy, Shanghai Jiao Tong University, Shanghai 200240, China}

	\begin{abstract}
Periodically driven systems provide unique opportunities to investigate the dynamics of topological excitations far from equilibrium. In this paper, we report a time-crystalline soliton (TCS) state in a driven-dissipative $\phi^4$ model. This state exhibits spontaneous breaking of discrete time-translational symmetry while simultaneously displaying spatial soliton behavior. During time evolution, the soliton pattern periodically oscillates between kink and anti-kink configurations. We further study TCS dynamics under noise, demonstrating that soliton random walk can induce a dynamical transition between two distinct $Z_2$ symmetry-breaking time-crystalline phases in time domain.  Finally, we examine the annihilation of two spatially separated TCSs under noise. Importantly, in contrast to the confined behavior of time-crystalline monopoles reported in [Phys. Rev. Lett. 131, 056502 (2023)], the dynamics of time-crystalline solitons is deconfined despite the nonequilibrium nature of our model: the statistically averaged annihilation time scales as a power law with the solitons' initial separation.

\end{abstract}

\maketitle 
\section{Introduction}
Driving systems out of equilibrium unlocks novel possibilities for spontaneous symmetry breaking (SSB) extending into the temporal domain. A paradigmatic example is the discrete time crystal (DTC) phase, which spontaneously breaks the discrete time-translational symmetry (DTTS) inherent to periodically driven Hamiltonians. DTCs have been the subject of intense experimental and theoretical investigation across diverse physical platforms in recent years\cite{choi2017observation,zhang2017observation,frey2022realization,autti2018observation,kessler2021observation,mi2022time,trager2021real,zhu2019dicke,gong2018discrete,yang2021dynamical,hu2023solvable,surace2019floquet,huang2018clean,russomanno2017floquet,munoz2022floquet,yao2017discrete,collado2021emergent,von2016absolute,Estarellas2020,yao2020classical,yue2022thermal,lupo2019nanoscopic,gambetta2019classical,khasseh2019many,hurtado2020building,ye2021floquet,pizzi2021classical}. Introducing spatial degrees of freedom further enriches this landscape, enabling the emergence of complex nonequilibrium phases characterized by intriguing spatiotemporal symmetry breaking\cite{li2012space,smits2018observation,fan2024emergence,yue2023space,kleiner2021space,trager2021real,PhysRevLett.131.221601,luo2024discrete}.

Beyond identifying new phases, a profound fundamental question arises: What are the elementary excitations residing above such far-from-equilibrium, symmetry-broken states? In equilibrium condensed matter, elementary excitation is closely related to spontaneous symmetry breaking.  Elementary excitations define the low-energy physics – phonons propagate through crystals, while topological excitations like domain walls (solitons) mediate transitions between degenerate ground states in Ising ferromagnets breaking discrete $Z_2$ symmetry. Extending this concept to highly excited, dissipative nonequilibrium phases remains a significant challenge and opportunity. Recent work has begun exploring topological excitations within DTC frameworks. Slow parameter ramping, for instance, can induce a dynamical transition between two distinct $Z_2$ symmetry-breaking DTC phases, interpreted as a temporal analog of a spatial soliton\cite{yang2021dynamical}. Furthermore, studies of monopole excitations in a time-crystalline spin ice revealed dynamical confinement – a direct consequence of the underlying nonequilibrium drive, contrasting sharply with the deconfined dynamics of their equilibrium counterparts\cite{yue2023prethermal}.

In this paper, we report the discovery and characterization of a fundamental time-crystalline soliton (TCS) state in a minimal driven-dissipative $\phi^4$ model. Crucially, dissipation plays a vital role by preventing runaway heating to a featureless infinite-temperature state, thereby stabilizing the essential DTC order. The TCS represents a localized topological defect exhibiting spontaneous DTTS breaking while simultaneously displaying characteristic solitonic structure in space; its profile periodically oscillates between kink and antikink configurations during one driving period. Strikingly, we demonstrate that: (i) Under the influence of white noise, the random walk of the TCS center dynamically drives transitions between two symmetry-broken time-crystalline phases, establishing a direct temporal counterpart to spatial soliton. (ii) More significantly, the dynamics of spatially separated TCSs is deconfined. Specifically, the statistically averaged annihilation time for a soliton-antisoliton pair exhibits a power-law scaling $\tau_{\text{ann}} \propto (\Delta L)^{\alpha}$ with their initial separation $\Delta L$. This stands in stark contrast to the exponentially confined dynamics ($\tau_{\text{ann}} \propto e^{\kappa \ell}$) observed for monopole pair in time-crystalline spin ice, where $\ell$ is the Dirac string length\cite{yue2023prethermal,zeng2023universal,suzuki2024topological}. 

\section{Model and method}

 We begin with reviewing the equilibrium properties of the one-dimensional single-component $\phi^4$ theory described by the free energy functional\cite{kardar2007statistical}: 
 
 \begin{align}
 F=\int dx \{\frac{1}{2}(\frac{d\phi}{dx})^2+V[\phi(x)]\}, 
 \end{align}
 
 where the potential $V[\phi(x)]=(\phi^4-2\epsilon\phi^2)/8$.  This model exhibits a discrete  $Z_2$ spin-flip symmetry: $\phi\to-\phi$. For $\epsilon>0$, the potential $V[\phi(x)]$ develops a double-well structure with degenerate symmetry-broken ground states at $\phi_\pm=\pm \sqrt{\epsilon}$.   Topological soliton connecting these two SSB states satisfy the Euler-Lagrange equation: $\frac{\delta H}{\delta \phi(x)}=\frac{\partial V}{\partial \phi}-\frac{d^2\phi}{dx^2}=0$, whose well-known kink solution takes the form  $\phi(x)=\sqrt{\epsilon} \mathrm{tanh}[\frac{\sqrt{\epsilon}(x-x_0)}{2}]$.

To investigate nonequilibrium phenomena, we consider the real-time evolution of a scalar field $\phi(x,t)$ coupled to a thermal bath. The dynamics obeys the stochastic Langevin equation\cite{laguna1997density}  

\begin{align}
    \frac{\partial^2\phi}{\partial t^2}+\eta \frac{\partial \phi}{\partial t}-\frac{\partial^2\phi}{\partial x^2}+\frac{\partial V}{\partial \phi}=\xi(x,t),\label{2}
\end{align}

Where $\eta $ is the dissipation strength fixed as $\eta=0.5$ throughout this paper. $\xi(x,t)$ is a
zero-mean $[\langle \xi(x,t)\rangle_\xi=0]$ random field representing thermal fluctuations. And the stochastic variables satisfy

 \begin{align}
     \langle \xi(x,t),\xi(x^\prime,t^\prime)\rangle_\xi=\mathcal{D}^2\delta(t-t^\prime)\delta(x-x^\prime),
 \end{align}

$\mathcal{D}$ is the strength of the noise, and the average $\langle \star\rangle_\xi$ is over all the noise trajectories. If the bath is in thermal equilibrium with temperature $\mathrm{T}$, the fluctuation dissipation theorem indicates that  $\mathcal{D}^2=2\eta \mathrm{T}$. 

 To engineer nonequilibrium temporal order, we introduce periodic modulation of the symmetry-breaking parameter: $\epsilon=J\mathrm{cos}(\omega_0 t)$. This drives the system between symmetric ($\epsilon<0$) and symmetry-broken ($\epsilon>0$) regimes at frequency $\omega_0$, creating conditions for spontaneous DTTS breaking.  The inertial mass term ($\partial_t^2\phi$)  enables emergent temporal order by preventing adiabatic following of the drive—its absence would force strict synchronization with the driving frequency $\omega_0$, rendering period-doubling impossible.  The dissipation in the model plays a crucial role in stabilizing the potential DTC order. It allows the system to relax towards a steady state where the energy injected by the periodic driving is dissipated away, preventing the system from heating up and maintaining it in a nonequilibrium state.  

Simulating stochastic partial differential equations (SPDEs) requires numerical methods that discretize both the spatial domain and the stochastic process.  The stochastic partial differential equation\ref{2} is discretized by adopting Stratonovich’s formula, and solving it by the standard Heun method with a time step of $\Delta t=10^{-2}$ and a space discretization with $\Delta x=0.1$, the convergence of which has been checked numerically (see Appendix B).  The system size in our simulation is up to $2L$  with $L=50$, which is more than enough for the simulation (see also Appendix B) . In the following, we will first focus on the long-time asymptotic dynamics of this model in the absence of noise and investigate the non-equilibrium  topological structure, i.e. TCS,  in spacetime. Then we will explore the random walk of TCS and the annihilation of a TCS pair in the presence of generic environmental noise.

\section{results and discussion}

\subsection{Time-crystalline solitons }
\begin{figure}
		\centering
		\includegraphics[width=0.5\textwidth]{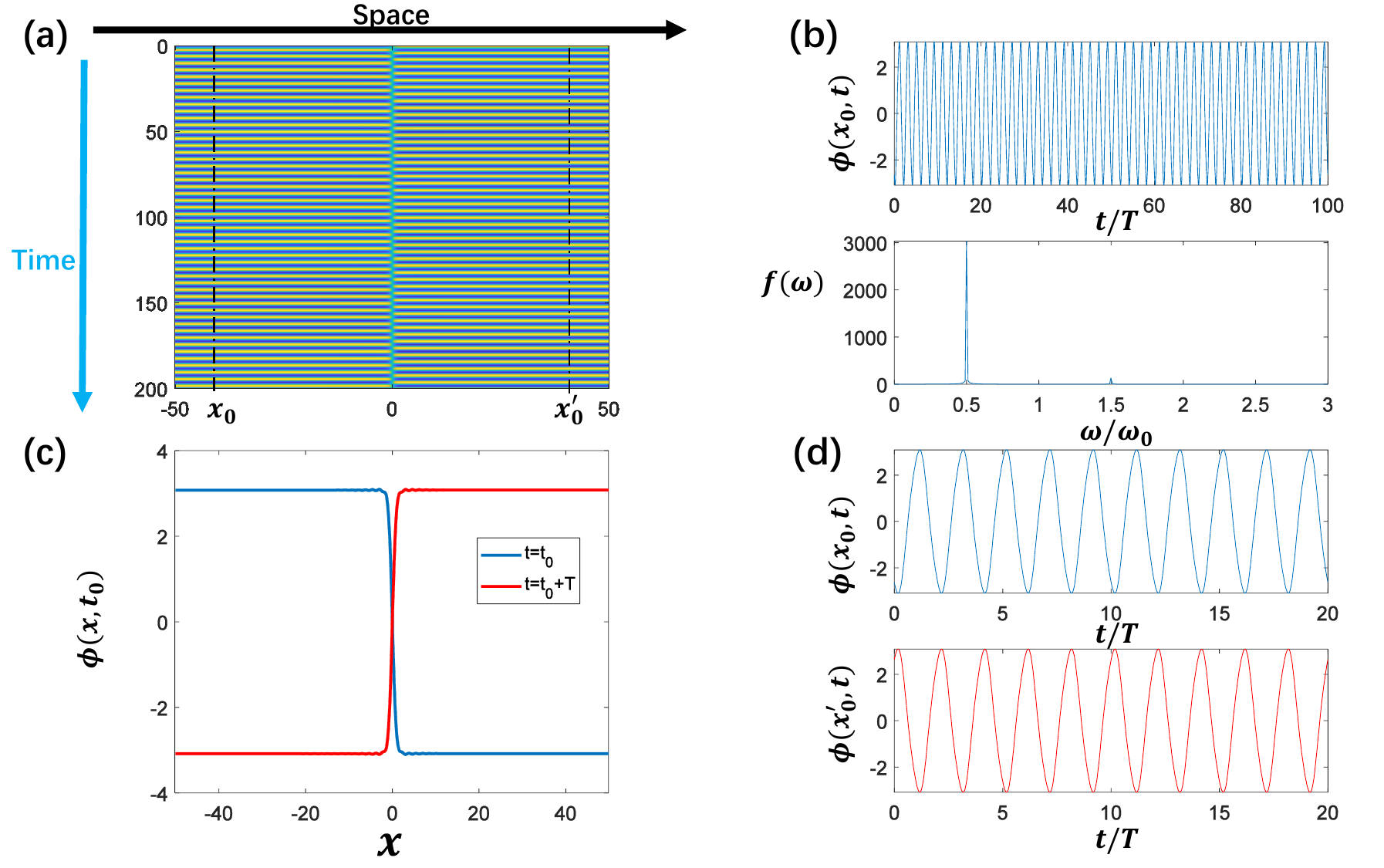}
		\caption{Time-crystalline soliton (TCS) in the driven-dissipative $\phi^4$ model. (a) Spatiotemporal evolution of $\phi(x,t)$. The bright central line traces the soliton center. (b)  Top: Period-doubled dynamics of $\phi(x_0,t)$ showing $2T$-periodicity ($T=2\pi/\omega_0$). Bottom: Fourier spectrum with dominant peak at $\omega_0/2$, confirming discrete time-translational symmetry breaking. (c) Snapshot of the TCS at time slice $t_0$ and $t_0+T$. (d) Two degenerate DTC phases related to each other by shifting $T$ along the temporal direction. $x_0=-40$, and $x_0^\prime=40$. Parameters in our simulation are chosen as $\eta=0.5$, $J=5$, $\omega_0=\pi$, $\mathcal{D}=0$.}\label{fig1}
\end{figure}

We start the time evolution with a topological soliton initial state, i.e. $\phi(x,0)=\sqrt{A} \mathrm{tanh}[\frac{\sqrt{A}(x-x_0)}{2}]$ under open boundary condition. We note that the long-time dynamics does not depend on the specific value of $A$. (See Appendix \ref{appendixB} for comparison with different initial states) . Then we monitor the evolution of the scalar field $\phi(x,t)$ under the periodic driving $\epsilon=J\mathrm{cos}(\omega_0 t)$ according to Equation  \ref{2}.  Figure \ref{fig1} (a) is a plot of  $\phi(x,t)$ in the two-dimensional spacetime, where the field configuration periodically oscillates in time.  And the soliton core (the bright central line) preserves its localized structure during the time evolution. In Figure\ref{fig1} (b), we can directly observe that $\phi(x_0,t)$ exhibits a persistent oscillation with a period of $2T$, i.e. twice that of the driving period ($T=2\pi/\omega_0$).  And the Fourier spectrum with dominant peak at $\omega_0/2$ confirms the periodic doubling dynamics. This periodic doubling phenomenon is a manifestation of spontaneous $Z_2$ discrete time translational symmetry breaking[$\phi(x,t)=\phi(x,t+2T)\neq \phi(x,t+T)$]. Figure\ref{fig1}(c) depicts snapshot of $\phi(x,t)$ at time slice $t_0$ and $t_0+T$.  The soliton profile exhibits kink-antikink oscillation during one period of driving and it returns to its original profile only after two periods of driving , i.e.,  $2T$, which actually reflects the SSB of DTTS.  In summary, TCS is a non-equilibrium topological localized structure with DTC order.

 Notably, the TCS acts as a dynamical domain wall separating two degenerate DTC phases. Each DTC phase exhibits robust $2T$-periodicity while being related to each other by shifting a driving period $T$ along the temporal direction. This temporal degeneracy emerges from $Z_2$ symmetry breaking of DTTS. Figure\ref{fig1}(d) dipicts time evolution of $\phi(x,t)$ at two symmetric positions  ($x_0^\prime=-x_0$ ).  It can be seen $\phi(x_0,t)$ and $\phi(x_0^\prime,t)$ can be connected to each other by shifting a driving period $T$, which can be considered as two $Z_2$ symmetry breaking “degenerate” DTC phases.  We demonstrate below that stochastic motion of the TCS dynamically mediates transitions between these two degenerate DTC phases in the temporal direction.

\begin{figure}
		\centering
		\includegraphics[width=0.5\textwidth]{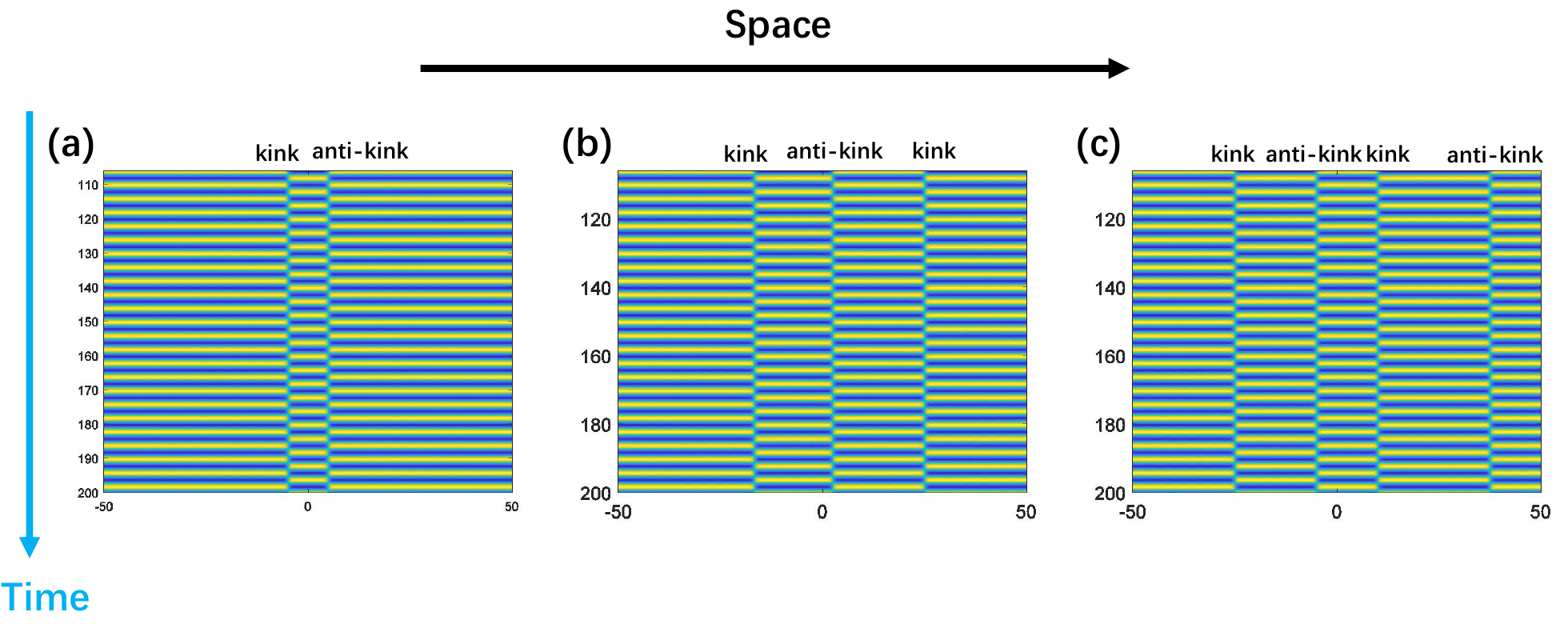}
		\caption{Stable multi-TCS configurations with kink and anti-kink.  Parameters in our simulation are chosen as $\eta=0.5$, $J=5$, $\omega_0=\pi$, $\mathcal{D}=0$.}\label{fig2}
	\end{figure}

Stable multi-TCS configurations can also exist in the driven-dissipative Landau-Ginzburg dynamics.  We initialize the system with a kink-antikink pair: $\phi(x,0) = A \left[ \tanh(k(x-x_1)) - \tanh(k(x-x_2)) \right] - A$ and then evolve under periodic driving . Figure\ref{fig2} (a) demonstrates robust time-crystalline bisoliton states. Similarly, Figure\ref{fig2} (b) and (c) reveal multisoliton solutions: (b) kink-antikink-kink trisoliton and (c) kink-antikink-kink-antikink tetrasoliton configurations, all exhibiting arbitrary spatial separation between TCSs. Again, all of these states break DTTS—manifested through subharmonic oscillations across all solitons—establishing them as fundamentally nonequilibrium phases inaccessible in equilibrium systems.

\subsection{Stochastic dynamics of TCS and its diffusive behavior}

    \begin{figure}
		\centering
		\includegraphics[width=0.5\textwidth]{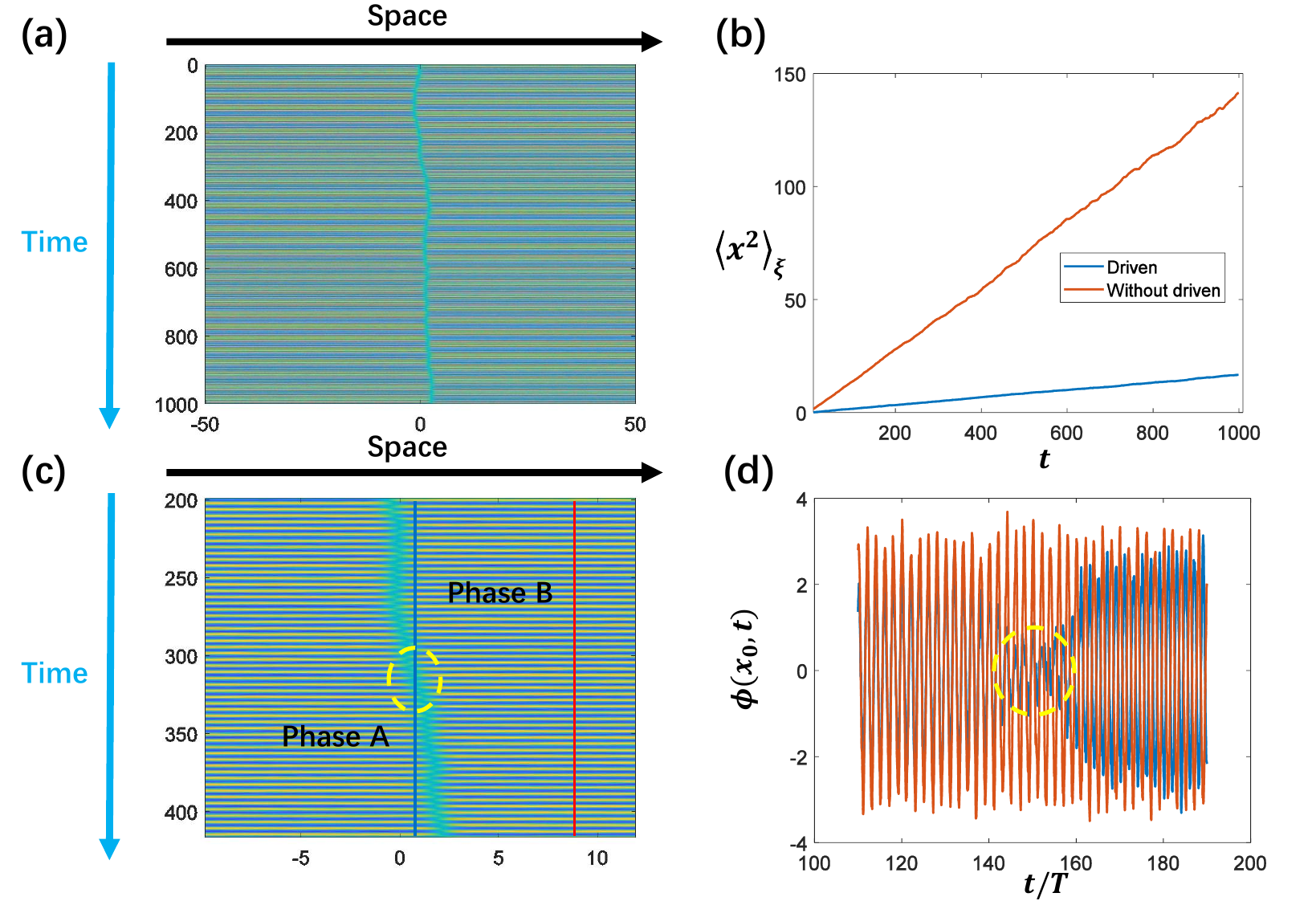}
		\caption{Random walk of TCS under noise and the intanton excitations.  (a) Dynamics of $\phi(x,t)$ under a single noise trajectory.  (b) Mean square displacement of the center of TCS compared to its equilibrium counterpart.  The ensemble average is taken over $2000$ noise trajectories.  (c)  Zoomed-in view of the region in (a). (d) Temporal evolution of $\phi(x,t)$ along the blue path (blue curve) and red path (red curve) indicated in (c).  The yellow circle indicates an instanton excitation between two degenerate $Z_2$ symmetry-breaking DTC phases.  Parameters in our simulation are chosen as $\eta=0.5$, $J=5$, $\omega_0=\pi$, $\mathcal{D}=0.5$.  }\label{fig3}
	\end{figure}

	\begin{figure}
		\centering
		\includegraphics[width=0.5\textwidth]{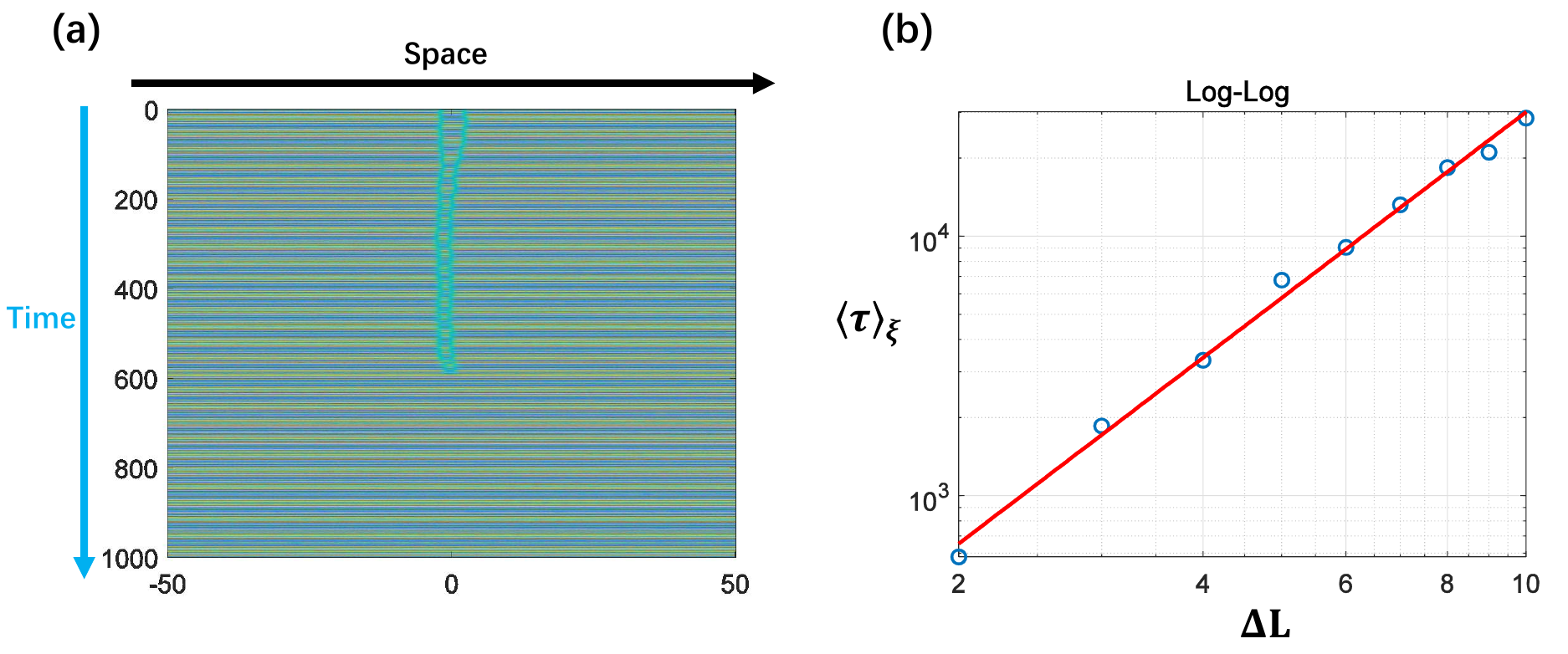}
		\caption{Annihilation of a kink-antikink pair. 
        (a) Evolution of the field $\phi(x,t)$ under a single noise trajectory. A kink and anti-kink annihilate after a characteristic time scale $\tau$. 
        (b) Average annihilation time $\langle \tau\rangle_\xi$ as a function of the initial separation between the solitons. The ensemble average is taken over 2000 noise trajectories. Other parameters are chosen the same as in Figure (\ref{fig3}).   }\label{fig4}
	\end{figure}


We now investigate the stochastic dynamics of the time-crystalline soliton under noise perturbations. Figure \ref{fig3}(a) displays the long-time spatiotemporal evolution of the TCS under a single noise  trajectory, revealing clear stochastic motion of the TCS center driven by thermal fluctuations. Quantitative analysis of the noise-induced diffusion is presented in Figure\ref{fig3}(b), where the ensemble-averaged mean squared displacement follows $\langle x(t)^2\rangle_\xi \sim t$, confirming normal diffusive behavior. The statistical ensemble average is taken over $2000$ independent noise trajectories. Figure \ref{fig3}(c) provides a magnified view of the temporal evolution from (a), while Figure\ref{fig3}(d) compares the field evolution $\phi(x_0,t)$ along two distinct paths indicated in (c). Crucially, the blue curve (crossing the soliton center) exhibits a $\pi$-phase shift, whereas the red curve (avoiding the core) shows no such phase inversion.  This $\pi$-phase shift serves as a dynamical transition between the two $Z_2$ symmetry-breaking degenerate DTC phases, and thereby constituting a topological defect in the temporal domain. In Figure\ref{fig3}(b), we observe that the TCS exhibits significantly reduced diffusion compared to equilibrium solitons in the undriven system. This suppression of diffusion arises because the activation of a DTC defect requires additional energy.

\subsection{Deconfinement of two spatially separated TCSs }

In equilibrium condensed matter systems, almost all excitations—such as spinons and holons in one-dimensional Luttinger liquids\cite{von1998bosonization}, half-charge solitons in conducting polymers\cite{heeger1988solitons}, and quasiparticles excitations in the fractional quantum Hall effect\cite{laughlin1999nobel}—are typically deconfined. Recent work, however, has revealed that nonequilibrium settings can fundamentally alter the deconfinement behavior. For example, in a two-dimensional time-crystalline spin ice, monopole excitations become dynamically confined despite their deconfined equilibrium counterparts\cite{yue2023prethermal}. There, a monopole-antimonopole pair is created by flipping a string of spins (the string is called Dirac string), and their average annihilation time scales as  $\tau_{\text{ann}} \propto e^{\kappa \ell}$ , where $\ell$ is the Dirac string length, signaling confinement.

Motivated by this, we investigate whether our time-crystalline solitons in 1D spatial dimension likewise experience confinement due to nonequilibrium driving. We initialize the system with a separated kink-antikink pair and track their stochastic dynamics under noise.  Figure\ref{fig4}(a) display a representative spatiotemporal evolution of the field $\phi(x,t)$ under a single noise trajectory, where the soliton and anti-soliton eventually annihilate each other. To determine whether the pair is confined, we compute the ensemble-averaged annihilation time $\langle \tau\rangle_\xi$ as a function of initial separation $\Delta L$ between the solitons.  The ensemble average is taken over 2000 noise trajectories. As shown in Figure\ref{fig4}(b), the statistically averaged annihilation time for a soliton-antisoliton pair exhibits a clear power-law scaling  $\langle \tau\rangle_\xi \propto (\Delta L)^{\alpha}$ with their initial separation $\Delta L$, indicating deconfined dynamics. This stands in sharp contrast to the exponential confinement observed in the 2D time-crystalline spin ice, highlighting that—even far from equilibrium—certain topological excitations can retain deconfinement.

\section{conclusion and outlook } 

In summary, by investigating a driven-dissipative $\phi^4$ model, we have uncovered a exotic time-crystalline soliton (TCS) state, which is a non-equilibrium topological localized structure that simultaneously breaks discrete time translational symmetry and exhibits spatial solitonic structure. The TCS manifests as a dynamical domain wall separating two degenerate $Z_2$ symmetry-breaking DTC phases, with its profile periodically oscillating between kink and antikink configurations during one driving period. Crucially, the stochastic motion of the TCS induces dynamical transitions between these two degenerate phases, realizing a topological excitation in the time domain. Furthermore, despite the strongly non-equilibrium nature of the system, the TCSs exhibit deconfined dynamics—the annihilation time of soliton-antisoliton pairs follows a power-law scaling with their initial separation, in sharp contrast to the exponentially confined dynamics observed in 2D non-equilibrium spin ice systems.



Our work opens several promising directions for future research. First, extending this framework to higher-dimensional systems may lead to the discovery of exotic non-equilibrium topological defects such as time-crystalline vortices in 2D or Skyrmions in 3D. A key open question is whether such objects remain deconfined—as observed for TCSs in our 1D setting—or become confined, as reported for monopoles in 2D time-crystalline spin ice in a driven-dissipative spin model. This raises a fundamental question: does spatial dimension play a decisive role in the confinement of non-equilibrium topological excitations? Second, it remains an exciting challenge to realize a quantum time-crystalline soliton or other exotic non-equilibrium quantum phases that simultaneously exhibit spatial topological order and long-range temporal order. In realistic quantum platforms, the interplay between quantum fluctuations and non-equilibrium topology is expected to give rise to rich, hitherto unexplored dynamical behavior.

 \section{ACKNOWLEDGMENTS}
We thank Zi Cai for useful discussions. We also thank Bo Fan for helpful comments on this work.  This work is supported by the National Key Research and Development Program of China (Grant No.2020YFA0309000), NSFC of China (Grant No.12174251), Natural Science Foundation of Shanghai (Grant No.22ZR142830), Shanghai Municipal Science and Technology Major Project (Grant No.2019SHZDZX01).

\appendix

\section{The algorithm for solving the stochastic Langevin equation}

 The stochastic Landau-Ginzburg equation \ref{2} can be solved by the standard Heun method. In every time step, we need to update the
field configuration $\phi(x,t)$. First, we decompose the second-order time derivative into two first-order equations:


\begin{align}
    \frac{\partial \phi}{\partial t}=y(x,t),\label{app1}
\end{align}
\begin{align}
    \frac{\partial y}{\partial t}=-\eta y+\frac{\partial^2\phi}{\partial x^2}-\frac{\partial V}{\partial \phi}+\xi(x,t),\label{app2}
\end{align}

Then we discretize spacetime as $x_i=i\Delta x$,  $   t_l=l\Delta t$,   where $i,l=0,1,2,\cdot\cdot\cdot$. Then the field configurations can be discretized as $\phi_{i}^l=\phi(x_i,t_l)$, $y_i^l= y(x_i,t_l)$.





 We calculate an intermediate configuration $\widetilde{\phi}_i^{l+1}$ as
\begin{align}
    \widetilde{\phi}_i^{l+1}=\phi_i^l+y_i^l\Delta t
\end{align}

Spatial derivatives are computed using the standard central difference scheme on both the current and  intermediate  configurations:

\begin{align}
    \frac{\partial^2\phi}{\partial x^2}=\frac{\phi_{i+1}^l-2\phi_i^l+\phi_{i-1}^l}{(\Delta x)^2}
\end{align}

\begin{align}
    \widetilde{\frac{\partial^2\phi}{\partial x^2}}=\frac{\widetilde{\phi}_{i+1}^{l+1}-2\widetilde{\phi}_i^{l+1}+\widetilde{\phi}_{i-1}^{l+1}}{(\Delta x)^2}
\end{align}

Equation (\ref{app2}) is discretized as:
\begin{widetext}
\begin{align}
    \frac{y_i^{l+1}-y_i^l}{\Delta t}=-\eta \frac{(y_i^l+y_i^{l+1})}{2}+\frac{1}{2}(\frac{\partial^2\phi}{\partial x^2}+\widetilde{\frac{\partial^2\phi}{\partial x^2}})-\frac{1}{2}(\phi_i^l)^3+\frac{1}{2}\epsilon\phi_i^l
    +\xi_i^l,
\end{align}

Solving for $y_i^{l+1}$ yields the update rule:

\begin{align}
    y_i^{l+1}&=y_i^l\frac{1-\frac{1}{2}\eta \Delta t}{1+\frac{1}{2}\eta \Delta t}+(\frac{\Delta t}{2(\Delta x)^2})\frac{1}{1+\frac{1}{2}\eta \Delta t}[(\phi_{i+1}^l-2\phi_i^l+\phi_{i-1}^l)+(\widetilde{\phi}_{i+1}^{l+1}-2\widetilde{\phi}_i^{l+1}+\widetilde{\phi}_{i-1}^{l+1})]\nonumber \\
    &-\frac{1}{2}\frac{\Delta t}{1+\frac{1}{2}\eta \Delta t}(\phi_i^l)^3+\frac{1}{2}\epsilon\frac{\Delta t}{1+\frac{1}{2}\eta \Delta t}\phi_i^l+\xi_i^l\frac{\Delta t}{1+\frac{1}{2}\eta \Delta t}
\end{align}
\end{widetext}
where $\xi_i^l$ is a stochastic noise field satisfying

 \begin{align}
    \xi_i^l=\frac{\mathcal{D}}{\sqrt{\Delta t\Delta x}}g_i^l
\end{align}

  where $g_i^l$ is a random number satisfying the Gaussian distribution with $\mathcal{N}(0,1)$: $\langle g_i^{l}\rangle_{\boldsymbol{\xi}}=0$, $\langle (g_i^{l})^2\rangle_{\boldsymbol{\xi}}=1$.

In the final step, we calculate 
the field configuration at $(l+1)$th time step,

\begin{align}
    \phi_i^{l+1}=\phi_i^l+(y_i^l+y_i^{l+1})\frac{\Delta t}{2},
\end{align}

\begin{figure}
	 	\centering
	 	\includegraphics[width=0.5\textwidth]{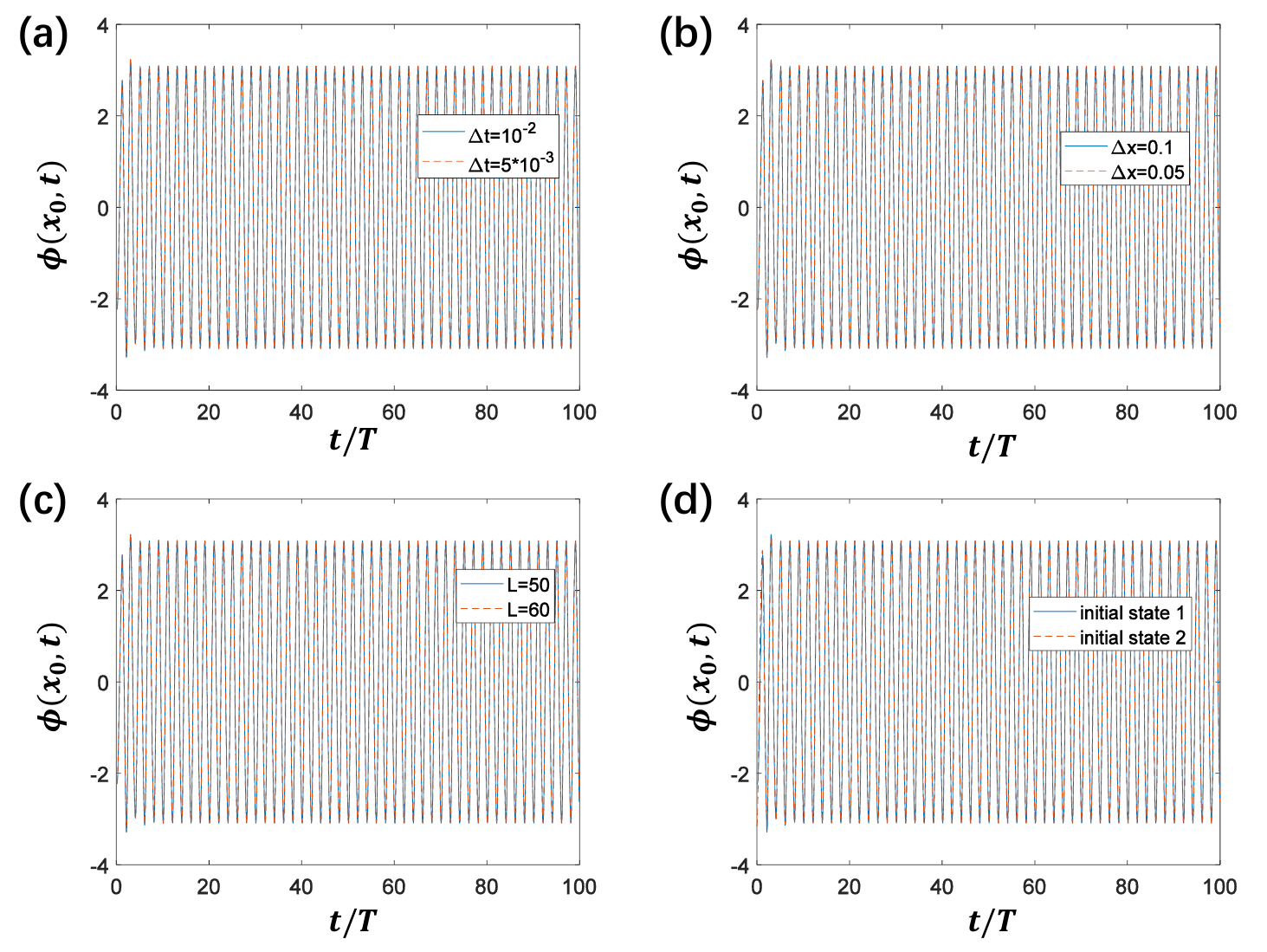}
	 	\caption{Comparison between dynamics with different $\Delta t$,  $\Delta x$,  $L$ and different initial states. Other parameters are chosen the same as Figure\ref{fig1} in the main text.}\label{appendix1}
	 \end{figure}

\section{$J$-$\omega_0$ phase diagram}
\begin{figure}
	 	\centering
	 	\includegraphics[width=0.4\textwidth]{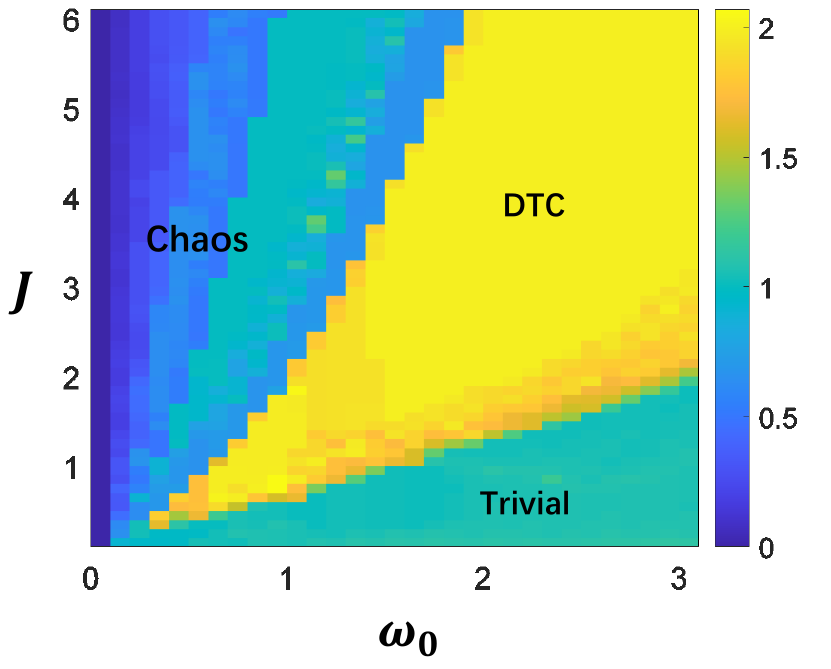}
	 	\caption{Zero temperature dynamical phase diagram in terms of the driving amplitude $J$ and the driving frequency $\omega_0$.}\label{appendix3}
	 \end{figure}

In this section, we explore the range of parameters $J$ and $\omega_0$ for which the DTC order occurs.  In summary, we plot the dynamical phase diagram in terms of the driving amplitude $J$ and the driving frequency $\omega_0$ in Fig.\ref{appendix3}, from which we can find that small driving amplitude favors a synchronized trivial phase, while large driving amplitude leads to an DTC phase.  We also notice that low driving frequency leads to more unstable chaotic behavior of the system.  The DTC order can be found for sufficiently large and fast driving.  Even though in the maintext, we focused only on a specific point in the parameter space $(J=5,\omega_0=\pi)$, the proposed TCS is a robust phase that exists in a large regime of the phase diagram.  This observation makes it easier to realize the TCS in a more realistic experimental setup.

\section{CONVERGENCE OF NUMERICAL RESULTS}\label{appendixB}

 \subsection{Finite $\Delta t$ and finite $\Delta x$} 
 
 In the main text, we choose a discrete time step of $\Delta t=10^{-2}$ and a space discretization of $\Delta x=0.1$. To check the convergence of our result with respect to these discretization parameters, we perform comparative simulations.  First we compare outcomes obtained using $\Delta t=10^{-2}$ and $\Delta t=5\times10^{-3}$.  As shown in Figure\ref{appendix1}(a), the results with different $\Delta t$  agree with each other quite well, which indicates that the $\Delta t=10^{-2}$ chosen in our simulation is sufficiently small, thus enabling us to ignore the errors induced by the discretization of time.  Similarly, we assess the spatial convergence by comparing results for   $\Delta x=0.1$ and $\Delta x=0.05$.  As shown in Figure\ref{appendix1}(b), the agreement between the two results indicates that $\Delta x=0.1$ chosen in our simulation is adequately fine to preclude significant errors from spatial discretization.

 \subsection{Finite $L$ } 

 Similarly, We would show that the finite system size $L=50$ used in the main text is enough for our simulations. Figure\ref{appendix1}(c) presents the comparison between dynamics with different $L$. The deviation between the results with $L = 50$ and $60$ is pretty small. Thus,  such a system size is sufficiently large that the finite-size effect can be neglected.

 \subsection{Initial states}
	 In the simulation of the main text, we begin with a soliton initial state,  i.e. $\phi(x,0)=\sqrt{A} \mathrm{tanh}[\frac{\sqrt{A}(x-x_0)}{2}]$ with $A=5$. Here, we demonstrate that the long-time dynamics does not depend on the specific value of $A$. We start the time evolution with different initial state with $A=5$ and $A=10$. Results in Figure\ref{appendix1}(d) show that the two results agree well with each other. This rapid loss of initial state memory can be attributed to the effects of dissipation.

	\bibliography{TCS.bib}

\end{document}